\documentclass[12pt,twoside,english]{article}
\usepackage[latin1]{inputenc}
\usepackage{pifont}
\usepackage{amssymb}

\makeatletter

\newcommand{\lyxaddress}[1]{
\par {\raggedright #1
\vspace{1.4em}
\noindent\par}
}

\AtBeginDocument{

}

\usepackage{babel}
\makeatother
\begin{document}

\title{ Composite Structure and Causality}

\author{Satish D. Joglekar\footnote{e-mail address:sdj@iitk.ac.in}}

\maketitle

\lyxaddress{Department of Physics, I.I.T. Kanpur, Kanpur 208016 (INDIA)}

\begin{abstract}

We study the question of whether a composite structure of elementary particles, with a length scale $1/\Lambda$, can leave observable effects of non-locality and causality violation at higher energies (but $\lesssim \Lambda$). We formulate a model-independent approach based on Bogoliubov-Shirkov formulation of causality. We analyze the relation between the fundamental theory (of finer constituents) and the derived theory (of composite particles). We assume that the fundamental theory is causal and formulate a condition which must be fulfilled for the derived theory to be causal. We analyze the condition and exhibit possibilities which fulfil and which violate the condition. We make comments on how causality violating amplitudes can arise.

\end{abstract}

\section{{ Introduction}}

The standard model (SM), a local quantum field theory, has served so far
as a very good description of elementary particle processes \cite{RPP}.
It is however widely believed that soon, when higher energies are
experimentally accessible, new phenomena may emerge that require a
description that goes beyond the standard model. Among the various
the possibilities, is the possibility that a composite nature of
the standard model constituents may be revealed \cite{comp} and
a possible failure of locality \cite{kh}. It is possible that the underlying
physics is nonlocal at shorter distances which could be a result of
composite structure of particles, or granularity of space-time, or
underlying noncommutative structure of space-time \cite{sz}. With
a nonlocal interaction, often goes causality violation that can arise
because the interaction region, encloses points separated by a space-like
interval. Causality violation has been studied in the context of non-local
\cite{jj04} and non-commutative quantum field theories \cite{CV,HJ06}.
It has in fact been suggested \cite{j01,jj04} that non-local quantum field theories \cite{ev91,kw92} may indeed serve as effective field theories for a deeper/more fundamental theory such as a composite model;  and the former indeed show causality violation\cite{kw92,jj04}. An effective tool to study causality has been developed by Bogoliubov
and Shirkov \cite{bs} and has been in particular employed for the
causality violation in non-local \cite{jj04} and non-commutative
QFT's \cite{HJ06}. We wish to consider the following question: in view of the possible composite
nature of elementary particles, leading to extended structures, will
these leave an observable effect in the form of a violation of causality
and locality that can be detected? A similar question regarding a
violation of the Pauli exclusion principle on account of the compositeness
of particles has been earlier addressed to \cite{PV}.  This question is particularly interesting
since should there be a signal of CV, it will be detected long before
an explicit knowledge of composite structure is known. In fact, it is has been suggested \cite{jj04} that the unknown physics at high energy scales ($\Lambda$) from a possible source can effectively be
represented in a consistent way (a unitary, gauge-invariant, finite (or renormalizable)
theory) by a nonlocal theory at energies lower than $\Lambda$, but higher than the present ones. In other words, the nonlocal standard model, with a parameter $\Lambda$, can
serve as such an effective field theory and will afford a model-independent way of consistently
reparametrizing the effects beyond standard model . In this model, one finds that there is but a small CV at low energies, which grows rapidly as energies approach $\Lambda$ and beyond these, the fundamental theory is expected to take over and presumably it leads to no CV again. The aim of the present work is to approach this question in a model-independent way in connection with a composite structure of SM constituents.

\section{{ Preliminary}}
\subsection{{ Definition of the problem}}

Suppose that the presently known standard model particles are a composite
of a set of finer constituents. Suppose that these underlying constituents
belong to a local causality-preserving \emph{fundamental} theory.
Suppose, further that at lower energies, one only observes the composite
bound states and their scattering processes. These bound state particles
are extended objects. A priori, their interaction is expected to be
non-local. A nonlocal covariant interaction has, at a given instant,
interaction spread over a region in space, which therefore contains
spatially separated points. An obvious question arises: will the interactions
of the \emph{composite} theory be such that causality is preserved
by this low-energy theory? We need the fundamental theory for energy
scales $>>\Lambda$, and for energy scales $<<\Lambda$, we have the
set of composite particles described by the "derived" theory. Then
the question, paraphrased differently is, will the \emph{phase transition}
(should there be one) from the fundamental to the composite be causality
preserving or it could lead to a breakdown of causality at short enough distances?

\subsection{{ Definition of the system}}

Let, for simplicity, the fundamental theory, denoted by $\mathcal{F}$,
be characterized by a single coupling constant \emph{$g$.} For the
purpose of formulation of the Bogoliubov-Shirkov (BS) criterion of
causality, we shall need to formulate a theory with a variable coupling
 $g(x)$. Let the low-energy derived theory be characterized
by its own coupling constant $\lambda\equiv\lambda\left[g\right]$,
which for identical reasons, we shall need to allow to depend on space-time:
$\lambda=\lambda\left(x\right)$. We shall assume, for simplicity,
that the derived theory, denoted by $\mathcal{C}$, is completely
described by its scattering states: \emph{i.e.} we shall assume that
the model admits no bound states. A scattering state of the derived
theory can be looked upon from \emph{two} different point of views:

\begin{itemize}
\item a scattering state, as $t\rightarrow\pm\infty$, is a state of a certain
set of non-interacting composite particles of the low energy theory
with certain momenta, polarizations etc.
\item a scattering state, as $t\rightarrow\pm\infty$, is a (complicated)
configuration of fields of the fundamental theory.
\end{itemize}

\section{{ Causality formulation for a theory without a well-defined S-matrix}}

Bogoliubov and Shirkov have shown \cite{bs} that $S-$operator is
causal \emph{only if} it satisfies,
\begin{equation}
\frac{\delta}{\delta g\left(x\right)}\left[\frac{\delta S}{\delta g\left(y\right)}S^{\dagger}\right]=0,\,\,\,\,\, x<\sim y\label{eq:causS}\end{equation}
(Here, $x<y$$\Longleftrightarrow x_{0}<y_{0}$ and $x\sim y\Longleftrightarrow(x-y)^{2}<0$
). The condition is obtained from the primary meaning of causality that a disturbance does not propagate outside the forward light-cone (the disturbance considered is that in  $g(x)$\footnote{One may consider varying $g(x)$ an unphysical operation, but one can look  alternately upon varying $g(x)$ at a point $x_0$ as insertion of a (specific) local operator $\frac{\partial\mathcal{L}_I}{\partial g(x)}$ at $x_0$ and study the propagation of its effects.}), and is independent of any specific field theory formulation. The BS causality criterion holds for a theory for which an S-operator
is defined. For a theory such as QCD, some of the matrix elements
of the S-operator may not exist on account of the infrared divergences.
It is nonetheless true that an alternate formulation in terms of the
U-operator (\emph{i.e.} $U\left(-T,T'\right)$) can be given. This
is so because, the U operator is unitary as much as the S-operator
and the BS criterion depends on two points $x,y$ with $x<\sim y$
which can always be chosen to be such that they both lie in $\left(-T,T'\right)$.
The relation would then read\begin{equation}
\frac{\delta}{\delta g\left(x\right)}\left[\frac{\delta U}{\delta g\left(y\right)}U^{\dagger}\right]=0,\,\,\,\,\, x<\sim y;\;\;\;-T<x_{0},y_{0}<T'\label{eq:causU}\end{equation}
It is possible to alternately formulate the causality condition in
terms of the following choices of the couplings. [This way results when we suitably integrate (\ref{eq:causU}) over $x_0<0$ and $y_0>0$]. In this approach, we make a comparison
of the following two \emph{neighboring} theories in the coupling constant
space%
\footnote{The idea of varying the coupling with time over all space is not an
entirely unfamiliar one: it is also employed in the LSZ formulation.%
}:

\begin{enumerate}
\item Fundamental theory $\mathcal{F}'$: Coupling constants = $g'_{2}$ (
a constant value) for $x_{0}>0$ and $g_{1}$(a constant value) for
$x_{0}<0$. Corresponding derived theory is $\mathcal{C}'$.
\item Fundamental theory $\mathcal{F}''$: Coupling constants = $g_{2}''$
( a constant value) for $x_{0}>0$ and $g_{1}$(the same constant
value) for $x_{0}<0$. Corresponding derived theory is $\mathcal{C}''$.
\end{enumerate}
All the coupling constants are (chosen to be) space-independent. It
suffices for our purpose that $g'_{2}$ differs infinitesimally from
$g_{1}$ and $g''_{2}$. (We can in fact assume that the infinitesimal change from $g_1$ to $g_2$ is carried out adiabatically and in an infinitesimal time). Then, we can alternately formulate \cite{sdj06} the
causality condition as,\begin{equation}
U\left[g_{1},g''_{2};-T,T'\right]U^{\dagger}\left[g_{1},g'_{2};-T,T'\right]\; is\; independent\; of\; g_{1}\label{eq:causA}\end{equation}
This alternate formulation makes mathematics simpler, though it may lead to an unusual-looking Physics. \\
In the following, we shall adopt a "reductio ad absurdum" approach: We shall let, if possible, that the theory $\mathcal{C}$ be causality-preserving and deduce the consequences of causality of $\mathcal{F}$ for $\mathcal{C}$ and analyze these.

\section{{ Relations between the derived theory and the fundamental
theory}}

\subsection{{ Relations between coupling constants}}

The coupling constant $\lambda$ is a function of $g$. If we allow
a space-time dependent coupling, then $\lambda=\lambda\left[g\right]$.
A small change%
\footnote{For the argument presented subsequently, we shall go back to a \emph{general} space-time
dependent coupling and not confine ourselves to the specific couplings presented in the previous section.}%
 $\delta g\left(x\right)$ in the coupling $g\left(x\right)$ about $g(x)=g=\mbox{constant}$, will
cause a change in $\lambda\left(x\right)$ as given by%
\footnote{{ We shall assume the existence and non-vanishing of $\left.\frac{\delta\lambda(z)}{\delta g(y)}\right|_{g(y)=g}$.
By translational invariance, this quantity is a function of $\left(z-y\right)$
and is independent of the point $z$ as such.}%
} $\delta\lambda(z)=\int dy\left.\frac{\delta\lambda(z)}{\delta g(y)}\right|_{g(y)=g}\delta g\left(y\right)$.
For the BS criterion of causality of Eq. (\ref{eq:causU}), we need to know the impact of a \emph{localized}
change $g(x)\rightarrow g\left(x\right)+\delta g\left(x\right)\equiv g\left(x\right)+\varepsilon\delta^{4}\left(x-\tilde{x}\right)$
on the function $\lambda\left(x\right)$. Now, if causality is valid,
$\lambda\left(x\right)$ cannot be affected for any $x_{0}<\tilde{x}_{0}$.
Assuming that the theory has T-invariance%
\footnote{We shall need that the theory $\mathcal{F}$ with a \emph{variable}
coupling has a T-invariance. This is possible to formulate a time-reversal
transformation for a theory with a variable $g\left(x\right)$: we
need to define the action of time-reversal as $Tg\left(\mathbf{x},t\right)T^{-1}=g\left(\mathbf{x,-}t\right)$.%
}\texttt{, $\lambda\left(x\right)$} cannot be affected for any $x_{0}>\tilde{x}_{0}$.
Thus, this \emph{together with causality} requires that,\begin{eqnarray*}
\lambda\left(x\right) & \rightarrow & \lambda\left(x\right)+C\varepsilon\delta^{4}\left(x-\tilde{x}\right)\\
 & + & terms\; having\; finite\; order\; derivatives\,\, of\,\, delta\,\, function\end{eqnarray*}
 Thus,\begin{eqnarray*}
\frac{\delta\lambda\left(z\right)}{\delta g\left(y\right)} & = & C\delta^{4}\left(z-y\right)\\
 & + & terms\; having\; finite\; order\; derivatives\,\, of\,\, delta\,\, function.\end{eqnarray*}
Then, for a \emph{constant small} change $\delta g=\varepsilon$,
for all  $x_{0}>0$, [i.e. $\delta g(x)=\epsilon\theta(x_0)$]; we find,\begin{eqnarray*}
\delta\lambda(z) & = & \int d^{4}y\frac{\delta\lambda(z)}{\delta g(y)}\delta g\left(y\right)\\
 & = & \int d^{4}y\left\{ C\delta^{4}\left(z-y\right)+derivatives\,\, of\,\, delta\,\, function\right\} \varepsilon\theta\left(y_{0}\right)\\
 & = & C'\varepsilon\theta\left(z_{0}\right)\;\; \mbox{for}\;{z}_{0}>0.\end{eqnarray*}
We shall denote by $\lambda'_{2}=\lambda[{g}_{2}',g_{1}]$ and $\lambda_{2}''=\lambda[{g}_{2}'',g_{1}]$.\\

\subsection{{ Relation between states}}

We shall work in the interaction picture of $\mathcal{C}$. Let the derived theory
$\mathcal{C}'$ have as incoming states%
\footnote{For technical simplicity, we shall assume that the set of states is
countably infinite.%
} $\left\{ \left|\tilde{c}_{m}\left(\lambda_{1},-T\right)\right\rangle \right\} $
which, as $-T\rightarrow-\infty$, represents scattering states with
a number of free composite particles. We shall keep $T$ finite and will
let $T\rightarrow\infty$ only at the end of the argument. Evidently,
as $-T\rightarrow-\infty$, $\left|\tilde{c}_{m}\left(\lambda_{1},-T\right)\right\rangle $
depends on $\lambda_1$ only through the self-interaction of each individual non-interacting particle in the state.
Let $\tilde{\mathcal{H}}$ denote the Hilbert space of states of $\mathcal{C}'$.
Then the hypothesis that the scattering states of $\mathcal{C}'$
forms a complete set implies that the set $\left\{ \left|\tilde{c}_{m}\left(\lambda_{1},-T\right)\right\rangle \right\} $
spans $\tilde{\mathcal{H}}$: $\tilde{\mathcal{H}}\equiv sp\left\{ \left|\tilde{c}_{m}\left(\lambda_{1},-T\right)\right\rangle \right\} $.
We shall denote by $\mathcal{H}$, the Hilbert space of states of
$\mathcal{F}'$ (and likewise for $\mathcal{F}''$ ).
Consider a state $\left|\tilde{c}_{m}\left(\lambda_{1},-T\right)\right\rangle \in  \tilde{\mathcal{H}}$
in the interaction picture. On physical grounds, we know that there is a corresponding state of $\mathcal{F}'$ in the interaction
picture, denoted by $\left|c_{m}\left(g_{1},-T\right)\right\rangle $. We note
that $\mathcal{H}$ can, in addition, have states linearly independent
of the states $\left\{ \left|c_{m}\left(g_{1},-T\right)\right\rangle \right\} $. We
augment this set to complete an orthonormal basis $\left\{ \left|c_{m}\left(g_{1},-T\right)\right\rangle \right\} \cup\left\{ \left|\beta_{n}\left(g_{1},-T\right)\right\rangle \right\} \equiv\left\{ \left|\alpha_{p}\left(g_{1},-T\right)\right\rangle \right\} $
for $\mathcal{H}$. We shall call the span of $\left\{ \left|c_{m}\left(g_{1},-T\right)\right\rangle \right\} $ by
$\hat{\mathcal{H}}\subset\mathcal{H}$.
A similar discussion holds for $\mathcal{F}''$. Let us now consider the time-evolution, from $t=-T$ to $t=T'$, of a single particle state of $\mathcal{C''}$ denoted by $\left|\tilde{s}_{p}\left(\lambda_{1},-T\right)\right\rangle$, which belongs to the basis of $\tilde{\mathcal{H}}$. The unitary time evolution operator  $\tilde{U}\left[\lambda_{1},\lambda''_{2};-T,T'\right]$ as applied to the state leads to
\begin{equation}
\tilde{U}\left[\lambda_{1},\lambda''_{2};-T,T'\right]\left|\tilde{s}_{p}\left(\lambda_{1},-T\right)\right\rangle=\left|\tilde{s}_{p}\left(\lambda''_{2},T'\right)\right\rangle\in \tilde{\mathcal{H}}
\end{equation}
This state is a single particle state of slightly different mass, on account of a slightly different self-energy, and interacts with a coupling $\lambda''_2$. We shall also introduce interaction picture states $\left|\tilde{d}_{m}\left(\lambda''_{2},T'\right)\right\rangle $. These states are at $t=T'$ and as $T'\rightarrow\infty$ consist of a set of non-interacting (but self-interacting) particles of a slightly different mass and coupling constant $\lambda''_2$. These are analogues of the "out" states. We shall assume that these also span $\tilde{\mathcal{H}}$. We shall further make a convention: Under time reversal, the quantum numbers of particles in the state $\left|\tilde{c}_{m}\left(\lambda_{1},-T\right)\right\rangle $ become those of $\left\langle \tilde{d}_{m}\left(\lambda_{1},T'\right)\right|$. Now, consider an exclusive process in $\mathcal{C}''$. The magnitude of the quantum mechanical amplitude for it, as seen
from $\mathcal{C}''$and $\mathcal{F}''$ are identical, as these are, in principle, experimentally observable:\begin{eqnarray}
 &  & |\tilde{u}_{nm}|\equiv \left|\left\langle \tilde{d}_{n}\left(\lambda''_{2},T'\right)\right|\tilde{U}\left[\lambda_{1},\lambda''_{2};-T,T'\right]\left|\tilde{c}_{m}\left(\lambda_{1},-T\right)\right\rangle \right| \nonumber \\
 & \equiv & \left|\left\langle d_{n}\left(g''_{2},T'\right)\right|U\left[g_{1},g''_{2};-T,T'\right]\left|c_{m}\left(g_{1},-T\right)\right\rangle \right| \equiv |u_{nm}|\label{eq:equi}
 \end{eqnarray}
Here, we have introduced states $\left|d_{n}\left(g''_{2},T'\right)\right\rangle$ in $\mathcal{H}$ analogous to $\left|\tilde{d}_{n}\left(\lambda''_{2},T'\right)\right\rangle$ in $\tilde{\mathcal{H}}$. We note that $U$ here is the $U-$matrix in the \emph{interaction}
picture of $\mathcal{F}''$, as the set of states $\left\{ \left|c_{m}\left(g_{1},-T\right)\right\rangle \right\} $
evolve according to the interaction Hamiltonian $\mathcal{H'}_{I}\left(g\right)$
(in the interaction picture) of the $\mathcal{F}''$.

First we note that on account of unitarity of $\tilde{U}$ and Eq. (\ref{eq:equi}),\begin{eqnarray}
1 & = & \sum_{n}\left|\left\langle \tilde{d}_{n}\left(\lambda''_{2},T'\right)\right|\tilde{U}\left[\lambda_{1},\lambda''_{2};-T,T'\right]\left|\tilde{c}_{m}\left(\lambda_{1},-T\right)\right\rangle \right|_{}^{2}\nonumber \\
 & = & \sum_{n}\left|\left\langle d_{n}\left(g''_{2},T'\right)\right|U\left[g_{1},g''_{2};-T,T'\right]\left|c_{m}\left(g_{1},-T\right)\right\rangle \right|^{2}\label{eq:uni1}\end{eqnarray}
and\begin{eqnarray}
1 & = & \sum_{m}\left|\left\langle \tilde{d}_{n}\left(\lambda''_{2},T'\right)\right|\tilde{U}\left[\lambda_{1},\lambda''_{2};-T,T'\right]\left|\tilde{c}_{m}\left(\lambda_{1},-T\right)\right\rangle \right|_{}^{2}\nonumber \\
 & = & \sum_{m}\left|\left\langle d_{n}\left(g''_{2},T'\right)\right|U\left[g_{1},g''_{2};-T,T'\right]\left|c_{m}\left(g_{1},-T\right)\right\rangle \right|^{2}\label{eq:uni2}\end{eqnarray}
So, the unitarity of $U$ implies, \[
\left\langle d_{n}\left(g''_{2},T'\right)\right|U\left[g_{1},g''_{2};-T,T'\right]\left|\beta_{m}\left(g_{1},-T\right)\right\rangle =0\]
\begin{equation}
\left\langle \beta_{n}\left(g''_{2},T'\right)\right|U\left[g_{1},g''_{2};-T,T'\right]\left|c_{m}\left(g_{1},-T\right)\right\rangle =0\label{eq:rest}\end{equation}
The relations (\ref{eq:rest}) implies that $U$ is a block-diagonal matrix. The unitarity of $ U$ then implies that the block corresponding to the subspace $\mathcal{\hat{H}}$, viz. $\hat{U}$, is also unitary. We shall now attempt relate these further. In this connection, we recall a result for a finite dimensional matrices:\\
\textbf{\emph{Lemma }}: Let $U$ and $U'$ be two $N\times N$ unitary matrices satisfying:
$|u'_{ij}|=|u_{ij}|;\;\; 1\leq i,j\leq N$. Then, there exist phases $\{\theta_i:i=1,2,\ldots,N\}$ and $\{\phi_i:i=2,\ldots,N\}$ such that $u'_{ij}=u_{ij}\exp{\left[i \left(\theta_i+\phi_j\right)\right]}:\; 1\leq i,j\leq N\;\;\mbox{with}\,\phi_1\equiv 0.$.\\
\textbf{\emph{Proof}}:
Let the diagonal elements of $U'$ and $U$ be related by: $u'_{ii}=\exp{(i\Theta_i)}u_{ii}$. We define $U''$ by $u''_{ij}=\exp{(-i\Theta_i)}u'_{ij}$. Then, $u''_{ii}=u_{ii}$. Now $U''$ is unitary and thus, a priori, has $N^2$ independent parameters. The information on moduli of elements constitutes $(N-1)^2$ independent conditions, corresponding to an $(N-1)\times(N-1)$ dimensional submatrix; the rest of the $(2N-1)$ moduli being determined by relations implying that the norm of each row and column is unity. The relations $u''_{ii}=u_{ii}$ imply \emph{additional} $N$ relations on the phases on $u_{ii}$.This leaves $N^2-(N-1)^2-N=N-1$ free parameters. The phases of $u''_{1j}, \;\; 2\leq j\leq N$ are unconstrained by $|u''_{ij}|=|u_{ij}|:\; 1\leq i,j\leq N$ and $u''_{ii}=u_{ii}$ and we define $u''_{1j}=u_{1j}\exp{(i\phi_j)}, \;\; 2\leq j\leq N$. Then, there are no free parameters and must lead to a unique $U''$. Now, $U''$ specified by $ u''_{ij}=u_{ij}\exp{(i\phi_j-i\phi_i)},\;\;1\leq i,j\leq N$ ($\phi_1\equiv 0$) is such a solution. This together with $u''_{ij}=\exp{(-i\Theta_i)}u'_{ij}$ leads to the result; with the definition $\Theta_i-\phi_i=\theta_i$.\\
\\
Thus,  in view of the unitarity of $\tilde{U}$, and $\hat{U}$ and (\ref{eq:equi}), we  write,
\begin{eqnarray}
 &  & \left\langle \tilde{d}_{n}\left(\lambda''_{2},T'\right)\right|\tilde{U}\left[\lambda_{1},\lambda''_{2};-T,T'\right]\left|\tilde{c}_{m}\left(\lambda_{1},-T\right)\right\rangle \nonumber \\
 & \equiv & \left\langle d_{n}\left(g''_{2},T'\right)\right|U\left[g_{1},g''_{2};-T,T'\right]\left|c_{m}\left(g_{1},-T\right)\right\rangle \times \exp{(i\theta''_n+i\phi_m)}\label{eq:equi1}
 \end{eqnarray}
 We shall assume that $\mathcal{F}$ and $\mathcal{C}$ are have time-reversal invariance and derive the consequences. Under time reversal, we know then that,
 \begin{equation}
\left\langle\beta\right|S\left|\alpha\right\rangle=\left\langle\mathcal{T}\alpha\right|S\left|\mathcal{T}\beta\right\rangle
\label{eq:TR} \end{equation}
where $\left|\mathcal{T}\beta\right\rangle$ is the state obtained by time-reversing the quantum numbers of the state $\left|\beta\right\rangle$. In this case, it would imply,
keeping in mind our choice of definitions,
\begin{eqnarray}
&  & \left\langle\tilde{d}_{n}\left(\lambda''_{2},T'\right)\right|\tilde{U}\left[\lambda_{1},\lambda''_{2};-T,T'\right]\left|\tilde{c}_{m}\left(\lambda_{1},-T\right)\right\rangle\nonumber  \\&  =&\left\langle\tilde{d}_{m}\left(\lambda_{1},T\right)\right|\tilde{U}\left[\lambda_{1},\lambda''_{2};-T,T'\right]\left|\tilde{c}_{n}\left(\lambda''_{2},-T'\right)\right\rangle \end{eqnarray}
We write a similar relation for $\mathcal{F}$. Putting $T'=T$, (or equivalently, noting that the matrix elements are insensitive to $T'$ and $T$), we find,
\begin{equation}
\phi_p(\lambda_2,\lambda_1)=\theta_p(\lambda_1,\lambda_2)\label{eq:trc}
\end{equation}

\section{Consequence of Causality of $\mathcal{F}$ for $\mathcal{C}$}
We shall assume that the fundamental theory $\mathcal{F}$ is causal and deduce the consequences for  the derived theory $\mathcal{C} (\mathcal{C'},\mathcal{C''})$. The causality of $\mathcal{F}$ implies that $$U\left[g_{1},g''_{2};-T,T'\right]U^{\dagger}\left[g_{1},g'{}_{2};-T,T'\right] $$ is independent of $g_1$. Hence, $$\mathcal{M}_{nm} \equiv \left\langle d_{n}\left(g''_{2},T'\right)\right|U\left[g_{1},g''_{2};-T,T'\right]U^{\dagger}\left[g_{1},g'{}_{2};-T,T'\right]\left|d_{m}\left(g'_{2},T'\right)\right\rangle $$
is also independent of $g_1$ since the state vectors $\left\langle d_{n}\left(g''_{2},T'\right)\right|$ and $\left|d_{m}\left(g'_{2},T'\right)\right\rangle $ are independent of $g_1$ with $g'_2$ and $g''_2$ fixed. We shall re-express $\mathcal{M}_{nm}$ in terms of the matrix elements of the derived theory $\mathcal{C} (\mathcal{C'},\mathcal{C''})$ and deduce the consequences. We note,
\begin{eqnarray}
\mathcal{M}_{nm}&  =&\left\langle d_{n}\left(g''_{2},T'\right)\right|U\left[g_{1},g''_{2};-T,T'\right]U^{\dagger}\left[g_{1},g'{}_{2};-T,T'\right]\left|d_{m}\left(g'_{2},T'\right)\right\rangle \nonumber  \\
& = &\sum_{p}\left\langle d_{n}\left(g''_{2},T'\right)\right|U\left[g_{1},g''_{2};-T,T'\right]\left|\alpha_{p}\left(g_{1},-T\right)\right\rangle \nonumber \\
 & \times & \left\langle \alpha_{p}\left(g_{1},-T\right)\right|U^{\dagger}\left[g_{1},g'_{2};-T,T'\right]\left|d_{m}\left(g'_{2},T'\right)\right\rangle \\
 &  =&\sum_{p}\left\langle d_{n}\left(g''_{2},T'\right)\right|U\left[g_{1},g''_{2};-T,T'\right]\left|c_{p}\left(g_{1},-T\right)\right\rangle \nonumber \\
 & \times & \left\langle c_{p}\left(g_{1},-T\right)\right|U^{\dagger}\left[g_{1},g'_{2};-T,T'\right]\left|d_{m}\left(g'_{2},T'\right)\right\rangle\\
 &  =&\sum_{p}\left\langle \tilde{d}_{n}\left(\lambda''_{2},T'\right)\right|\tilde{U}\left[\lambda_{1},\lambda''_{2};-T,T'\right]\left|\tilde{c}_{p}\left(\lambda_{1},-T\right)\right\rangle \exp{[-i(\tilde{\theta}''_p-\tilde{\theta}'_p)]}\nonumber \\
 & \times & \left\langle \tilde{c}_{p}\left(\lambda_{1},-T\right)\right|\tilde{U}^{\dagger}\left[\lambda_{1},\lambda'{}_{2};-T,T'\right]\left|\tilde{d}_{m}\left(\lambda'_{2},T'\right)\right\rangle \exp{-[i(\theta''_n-\theta'_m)]}\label{eq:MinC}\\
 &  \equiv  &\tilde{\mathcal{M}}_{nm}(\lambda''_2,\lambda'_2,\lambda_1)
\end{eqnarray}
In the $3^{rd}$ step, we have employed the equations (\ref{eq:rest}) and
in the second step, we have employed the closure relation for $\mathcal{F}$. \\ In the above, $\theta''_n\equiv \theta_n(\lambda''_2,\lambda_1)$,  $\theta'_m\equiv \theta_m(\lambda'_2,\lambda_1)$, and $\tilde{\theta}''_p\equiv \theta_p(\lambda_1,\lambda''_2)$ etc.
Thus, the expression (\ref{eq:MinC}) is independent of $\lambda_1$:
\begin{equation}
\frac{\partial\tilde{\mathcal{M}}_{nm}(\lambda''_2,\lambda'_2,\lambda_1)}{\partial\lambda_1}=0
\label{eq:CC}\end{equation}

\section{Analysis of Causality Condition}
We shall now analyze the condition (\ref{eq:CC}) obtained as an implication of causality of  $\mathcal{F}$.  For this purpose, we shall find it useful to Taylor-expand $\theta_n$ as follows\footnote{Throughout, we have employed only the infinitesimal variations in the couplings. These are  sufficient to determine the first order partial derivatives with respect to each $\lambda_1$ and $\lambda_2$. Hence, we shall content ourselves with expansion only upto $O(\Delta_1\Delta_2)$}:
\begin{eqnarray}
\theta_n(\lambda''_2,\lambda_1)& =&\theta_n(\lambda'_2,\lambda_{1(0)})+\beta_n\Delta_1+\gamma_n\Delta_2+\delta_n\Delta_1\Delta_2+\cdots\nonumber  \\& \equiv  &\alpha_n+\beta_n\Delta_1+\gamma_n\Delta_2+\delta_n\Delta_1\Delta_2+\cdots\nonumber  \\ \theta_m(\lambda'_2,\lambda_1)& =&\alpha_m+\beta_m\Delta_1+\cdots\label{eq:exp}
\end{eqnarray}
Here,
$\Delta_1\equiv \lambda_1-\lambda_{1(0)};\;\;\Delta_2\equiv \lambda''_2-\lambda'_2$;
 $\beta,\gamma, \delta$ refer to appropriate partial derivatives at $(\lambda'_2,\lambda_{1(0)})$ and $\lambda_{1(0)}$ is some value near $\lambda_1$.\\
We note that if \\
$$\theta_n(\lambda_2,\lambda_1)\; \mbox{is a function only of its \emph{first} argument }\qquad\qquad\mbox{(I)}\,$$\\
 then,
$(\tilde{\theta}''_p-\tilde{\theta}'_p)\equiv \theta_p(\lambda_1,\lambda''_2)-\theta_p(\lambda_1,\lambda'_2)$ is zero and $(\theta''_n-\theta'_m)\equiv \theta_n(\lambda''_2,\lambda_1)-\theta_m(\lambda'_2,\lambda_1)$ is independent of $\lambda_1$. Also, we can then carry out the sum over $p$ using the completeness relation and find that the independence from  $\lambda_1$ of
\begin{eqnarray}
&  &\tilde{\mathcal{M}}_{nm}(\lambda''_2,\lambda'_2,\lambda_1)\nonumber  \\
&=  &\left\langle \tilde{d}_{n}\left(\lambda''_{2},T'\right)\right|\tilde{U}\left[\lambda_{1},\lambda''_{2};-T,T'\right]\tilde{U}^{\dagger}\left[\lambda_{1},\lambda'{}_{2};-T,T'\right]\left|\tilde{d}_{m}\left(\lambda'_{2},T'\right)\right\rangle \nonumber  \\
&  \times & \exp{-[i(\theta''_n-\theta'_m)]}
\end{eqnarray}
for all $m,n$ implies $\tilde{U}\left[\lambda_{1},\lambda''_{2};-T,T'\right]\tilde{U}^{\dagger}\left[\lambda_{1},\lambda'{}_{2};-T,T'\right]$
is independent of $\lambda_1$. This condition is indeed necessary for causality of $\mathcal{C}$. In fact, in \emph{this} case, we can rewrite\footnote{We have dropped primes on $\lambda_2$.}
\begin{eqnarray}
 &  & \left\langle \tilde{d}_{n}\left(\lambda_{2},T'\right)\right|\tilde{U}\left[\lambda_{1},\lambda_{2};-T,T'\right]\left|\tilde{c}_{m}\left(\lambda_{1},-T\right)\right\rangle \nonumber \\
 & \equiv & \left\langle d_{n}\left(g_{2},T'\right)\right|U\left[g_{1},g_{2};-T,T'\right]\left|c_{m}\left(g_{1},-T\right)\right\rangle\nonumber  \\
 &  \times &   \exp{(i\theta_n(\lambda_2,\lambda_1)+i\theta_m(\lambda_1,\lambda_2))}\end{eqnarray}
as,
\begin{eqnarray}
 &  & \left\langle \tilde{d^*}_{n}\left(\lambda_{2},T'\right)\right|\tilde{U}\left[\lambda_{1},\lambda_{2};-T,T'\right]\left|\tilde{c^*}_{m}\left(\lambda_{1},-T\right)\right\rangle \nonumber \\
 & \equiv & \left\langle d_{n}\left(g_{2},T'\right)\right|U\left[g_{1},g_{2};-T,T'\right]\left|c_{m}\left(g_{1},-T\right)\right\rangle \end{eqnarray}
by redefining states by absorbing phases:$$(\left|\tilde{c^*}_{m}\left(\lambda_{1},-T\right)\right\rangle =e^{-i\theta_m(\lambda_1)}\left|\tilde{c}_{m}\left(\lambda_{1},-T\right)\right\rangle )$$ etc. We note that this redefinition of the states is meaningful and \emph{compatible with causality} when $\theta_n$ is independent of its second argument. If on the other hand, $\theta_n$ is dependent on its second argument (excepting a possibility below), we cannot absorb a phase in a manner \emph{compatible with causality}: a state $\left|\tilde{c^*}_{m}\right\rangle$ at $t= -T$ cannot be made to depend on the value of coupling $ \lambda_2$ it would have at a later time $t>0$.

We can, in fact, liberalize somewhat the above condition by requiring that,
$$\beta_n=\beta \;\;\qquad\mbox{and}\,\,\,\delta_n=0\qquad\forall \;\;n\qquad\qquad\mbox{(II)}\,$$
In this case,
\begin{eqnarray}
(\tilde{\theta}''_p-\tilde{\theta}'_p)&  \equiv &  \theta_p(\lambda_1,\lambda''_2)-\theta_p(\lambda_1,\lambda'_2)\nonumber  \\
&  =&\beta\Delta_2+\cdots
\end{eqnarray}
is independent of $\lambda_1$ and  does not depend also on $p$ and thus comes out of the summation in (\ref{eq:MinC}). The summation in (\ref{eq:MinC}) can be carried out using the completeness relation. Also, $(\theta''_n-\theta'_m)\equiv \theta_n(\lambda''_2,\lambda_1)-\theta_m(\lambda'_2,\lambda_1)$ is still independent of $\lambda_1$. Thus, the entire discussion proceeds as before: in particular, as a little analysis shows, the phases can again be absorbed into the definition of states in a manner compatible with causality.

While we shall not provide the general analysis of (\ref{eq:CC}), we shall establish examples of a few specific \emph{sufficient} conditions for causality violation. (These are simple conditions that, in fact, contradict $I$ or $II$ above) We can easily verify the following results:
\begin{enumerate}
  \item There is causality violation if (i) for some $n$, $\delta_n\neq 0$ and (ii) $\beta_n=\beta_m \quad\forall\;\; n,m$
  \item There is causality violation if  there be $m \neq n$  such that\\ $\tilde{\mathcal{M}}_{nm}(\lambda''_2,\lambda'_2,\lambda_1)\neq 0$, when evaluated to $O(\Delta)$, and $\beta_m \neq \beta_n$.
 \end{enumerate}
  \textbf{Proof}: We shall let, if possible, $\mathcal{C}$ be causal. We can then write,
\begin{equation}
\tilde{U}\left[\lambda_{1},\lambda''_{2};-T,T'\right]=\tilde{U}\left[\lambda''_{2};0,T'\right]\tilde{U}\left[\lambda_{1};-T,0\right]
\end{equation}
Then, we can write the expression (\ref{eq:MinC}) as,
\begin{eqnarray}
\tilde{\mathcal{M}}_{nm}(\lambda''_2,\lambda'_2,\lambda_1)& = &\sum_{p}\left\langle \tilde{d}_{n}\left(\lambda''_{2},0\right)\right|\left.\tilde{c}_{p}\left(\lambda_{1},0\right)\right\rangle \exp{[-i(\tilde{\theta}''_p-\tilde{\theta}'_p)]}\nonumber \\
 & \times & \left\langle \tilde{c}_{p}\left(\lambda_{1},0\right)\right.\left|\tilde{d}_{m}\left(\lambda'_{2},0\right)\right\rangle \exp{-[i(\theta''_n-\theta'_m)]}\nonumber  \\
 & \equiv  &\left\langle \tilde{d}_{n}\left(\lambda''_{2},0\right)\right|\mathcal{X}\left|\tilde{d}_{m}\left(\lambda'_{2},0\right)\right\rangle \exp{[-i(\theta''_n-\theta'_m)]}\label{eq:M'}
\end{eqnarray}
where $\mathcal{X}\equiv \sum_{p}\left|\tilde{c}_{p}\left(\lambda_{1},0\right)\right\rangle \left\langle \tilde{c}_{p}\left(\lambda_{1},0\right)\right|\exp{[-i(\tilde{\theta}''_p-\tilde{\theta}'_p)]}$. We shall now expand the quantities involved to the first order in the infinitesimals as in (\ref{eq:exp}). In addition, we note that to the zeroth order in $\Delta_2$, (i.e. $\lambda"_2-\lambda'_2=0$), we have,
$(\tilde{\theta}''_p-\tilde{\theta}'_p)=0$ and the completeness relation leads to $\mathcal{X}=1$. We further define,
\begin{equation}
 \left\langle \tilde{d}_{n}\left(\lambda''_{2},0\right)\right|\left.\tilde{d}_{m}\left(\lambda'_{2},0\right)\right\rangle =\delta_{nm}+i\eta_{nm}\Delta_2+\cdots \end{equation}
\emph{\textbf{Proof of (i)}}: We define $\delta_{0}\equiv max\{|\delta_n|\}$; and let  $ \pm\delta_0= \delta_q$ for some $q$. We now have,
\begin{eqnarray}
(\tilde{\theta}''_p-\tilde{\theta}'_p)&  \equiv &  \theta_p(\lambda_1,\lambda''_2)-\theta_p(\lambda_1,\lambda'_2)\nonumber  \\
&  =&\beta\Delta_2+\delta_p\Delta_1\Delta_2+\cdots
\end{eqnarray}
and thus,
\begin{eqnarray}
\mathcal{X}&  \equiv &  \sum_{p}\left|\tilde{c}_{p}\left(\lambda_{1},0\right)\right\rangle \left\langle \tilde{c}_{p}\left(\lambda_{1},0\right)\right|\exp{[-i(\tilde{\theta}''_p-\tilde{\theta}'_p)]}\nonumber  \\
& = &\exp{(-i\beta\Delta_2)}\sum_{p}\left|\tilde{c}_{p}\left(\lambda_{1},0\right)\right\rangle \left\langle \tilde{c}_{p}\left(\lambda_{1},0\right)\right|\times \exp{[-i(\delta_p\Delta_1\Delta_2)]}\nonumber  \\
& = &\exp{(-i\beta\Delta_2)}\left[\mathcal{I}-i\Delta_1\Delta_2\sum_{p}\left|\tilde{c}_{p}\left(\lambda_{1},0\right)\right\rangle \left\langle \tilde{c}_{p}\left(\lambda_{1},0\right)\right|\delta_p\right]
\end{eqnarray}
Thus,
\begin{eqnarray}
&  &\exp{(i\beta\Delta_2)}\left\langle \tilde{d}_{q}\left(\lambda''_{2},0\right)\right|\mathcal{X}\left|\tilde{d}_{q}\left(\lambda'_{2},0\right)\right\rangle\nonumber  \\&  =&1+i\eta_{qq}\Delta_2-i\Delta_1\Delta_2\sum_{p}\delta_p |u_{pq}|^2+\cdots
\end{eqnarray}
where $u_{pq}\equiv \left\langle \tilde{d}_{q}\left(\lambda_{2},0\right) \right| \left.\tilde{c}_{p}\left(\lambda_{1},0\right)\right\rangle$ (We can ignore primes on $\lambda_2$ in this term).
The multiplicative exponential factor in (\ref{eq:M'}) becomes:
$$\exp{(-i\gamma_q\Delta_2-i\delta_q\Delta_1\Delta_2+\cdots)}\approx1-i\gamma_q\Delta_2-i\delta_q\Delta_1\Delta_2+\cdots$$. Thus,
\begin{eqnarray}
\tilde{\mathcal{M}}_{qq}&=&1+i\eta_{qq}\Delta_2-i\gamma_q\Delta_2-i\delta_q\Delta_1\Delta_2-i\Delta_1\Delta_2\sum_{p}\delta_p |u_{pq}|^2+\cdots\nonumber  \\
& = &1+i\eta_{qq}\Delta_2-i\gamma_q\Delta_2-i\Delta_1\Delta_2\sum_{p}[\delta_q+\delta_p ] |u_{pq}|^2+\cdots
\end{eqnarray}
In view of the fact that either $\delta_p+\delta_q\geq 0\;\forall\;p$ or $\delta_p+\delta_q\leq 0\;\forall\;p$ the last term is necessarily non-vanishing and dependent on $\Delta_1$ \footnote{There is the obvious exception that $\delta_p= -\delta_q\;\;\mbox{for every such p such that}\,u_{pq}\neq 0$; and this has to be valid for each such q for which $\delta_q=\pm \delta_0$.}.\\
\emph{\textbf{Proof of (ii)}}: Consider the matrix element
\begin{eqnarray}
\tilde{\mathcal{M}}_{nm}(\lambda''_2,\lambda'_2,\lambda_1) & \equiv  &\left\langle \tilde{d}_{n}\left(\lambda''_{2},0\right)\right|\mathcal{X}\left|\tilde{d}_{m}\left(\lambda'_{2},0\right)\right\rangle \exp{-[i(\theta''_n-\theta'_m)]} \neq 0
\end{eqnarray}
for $n\neq m$. To the first order in the infinitesimals, the nonzero matrix element
$$ \left\langle \tilde{d}_{n}\left(\lambda''_{2},0\right)\right|\mathcal{X}\left|\tilde{d}_{m}\left(\lambda'_{2},0\right)\right\rangle $$
is independent of $\Delta_1$. The multiplicative phase factor,
$$\exp{-[i(\theta''_n-\theta'_m)]}=\exp\{-i(\alpha_n-\alpha_m)-i(\beta_n-\beta_m)\Delta_1-i\gamma_n\Delta_2\}$$
is necessarily dependent on $\Delta_1$, thus implying causality violation.

\section{Additional comments}

We comment in a qualitative way upon how a phase factor depending on both values of the coupling can arise. Suppose that the derived theory $\mathcal{C}$ is actually correctly described by a nonlocal covariant theory with a finite non-zero non-locality scale $\Delta \sim 1/\Lambda$. Since the theory is covariant,
it is also non-local in time.  We write,
\begin{eqnarray}
\tilde{U}(\lambda_1,\lambda_2;-T,T')&  =&\tilde{U}(\lambda_2;\Delta,T')\tilde{U}(\lambda_1,\lambda_2;-\Delta,\Delta)\tilde{U}(\lambda_1;-T,-\Delta)
\end{eqnarray}
where the first and the third factors on the right hand side depends only on one value of the coupling due to finite size of non-locality in time. The second factor however depends on both couplings because in this time-interval $(-\Delta,\Delta)$, time evolution depends on both values of the coupling $\lambda$. On the other hand, the fundamental theory, being local and causal, however has no such analogue . The matrix $\tilde{U}(\lambda_1,\lambda_2;-\Delta,\Delta)$ can then give rise to phases depending on both couplings in relation (\ref{eq:equi1}).\\
\\
Naively, one may expect that if the fundamental theory is causal, the derived theory should be so.
Examples are however known where the diagrams of the fundamental theory are associated with a different weight in the actual phenomenology. For example,  OZI rule in hadronic phenomenology gives a suppression of a subset of the QCD diagrams. While such a possibility is distinct from what is discussed in this work, generally such a modification of the amplitudes within the fundamental theory may alter the underlying properties of the fundamental theory such as causality.

\end{document}